\documentclass[aps,prb,twocolumn,showpacs,superscriptaddress,floatfix]{revtex4-1}

\pdfoutput=1
\usepackage[titletoc,toc,title]{appendix}
\usepackage{amsmath,amssymb}
\usepackage{graphicx}
\usepackage{color}
\usepackage{rotating}
\usepackage{bm}
\usepackage{setspace}
\usepackage{hyperref}
\usepackage{float}
\usepackage{soul}
\usepackage{color}
\usepackage{braket}

\hypersetup{
colorlinks=true,
urlcolor= blue,
citecolor=blue,
linkcolor= blue,
bookmarks=true,
bookmarksopen=false,
}

\begin{document}

\preprint{APS/123-QED}

\title{An exact continuum model for low-energy electronic states of twisted bilayer graphene}
\author{Stephen Carr}
\affiliation{Department of Physics, Harvard University, Cambridge, Massachusetts 02138, USA}
\author{Shiang Fang}
\affiliation{Department of Physics, Harvard University, Cambridge, Massachusetts 02138, USA}
\author{Ziyan Zhu}
\affiliation{Department of Physics, Harvard University, Cambridge, Massachusetts 02138, USA}
\author{Efthimios Kaxiras}
\affiliation{Department of Physics, Harvard University, Cambridge, Massachusetts 02138, USA}
\affiliation{John A. Paulson School of Engineering and Applied Sciences, Harvard University, Cambridge, Massachusetts 02138, USA}

\begin{abstract}
We introduce a complete physical model for the single-particle electronic structure of twisted bilayer graphene (tBLG), 
which incorporates the crucial role of lattice relaxation.
Our model, based on $k \cdot p$ perturbation theory, 
combines the accuracy of DFT calculations through 
effective tight-binding Hamiltonians
with the computational efficiency and complete control of the twist angle 
offered by continuum models.
The inclusion of relaxation significantly changes the bandstructure at the 
first magic-angle twist corresponding to flat bands near the Fermi level 
(the ``low-energy'' states), 
and eliminates the appearance of a second magic-angle twist.
We show that minimal models for the low-energy states of tBLG can be easily modified to capture the changes in electronic states as a function of twist angle. 
\end{abstract}

\maketitle

The discovery of correlated phases in
twisted bilayer graphene (tBLG) 
has generated much interest in this structurally and compositionally rather simple system; 
it has emerged as a new platform for tunable electronic correlations, and for 
exploring of the nature of unconventional superconductivity\cite{Cao2018mott, Cao2018sc}.
The challenge in modeling these phenomena from an atomistic perspective 
is that the actual structure of tBLG near the magic-angle twist ($\sim 1.1^\circ$) where 
correlated behavior is observed, consists of a large number of atoms, exceeding $10^4$.
To make progress from the theoretical point of view, a minimal model is needed that 
can capture the essence of single-particle states near the Fermi level 
(``low-energy'' states).
Such a model should reproduce the energy spectrum as a function of their relative twist angle with reasonable accuracy and with the required fidelity in capturing 
the {\em nature} of low-energy states.
The appearance of correlated behavior is related to bands
with very low dispersion (``flat'' bands)
caused by interlayer hybridization between the two Dirac cones from the different 
layers\cite{Li2009, Brihuega2012, Luican2011, Wong2015}.

Existing models based on DFT calculations \cite{Trambly2010, Uchida2014} 
or large supercell tight-binding Hamiltonians \cite{Morell2010, Nam2017, Angeli2018}
are too complex to form the basis of a realistic many-body theory.
At the other extreme, simplified 
continuum models allow for efficient calculations, but are based on heuristic 
arguments about the nature of the relevant 
electronic states \cite{Mele2010,Bistritzer2011,San-Jose2012, Weckbecker2016}.
An important feature of the physical system is the presence of atomic
relaxation near the magic-angle twist,
which has significant effects on the low-energy 
bandstructure \cite{Yoo2018, Dai2016, Nam2017, Zhang2018, Lin2018}.
Many simplified models for the flat bands of magic-angle tBLG have been 
proposed based on symmetry analysis, 
but they rely on empirical parameterization 
and are designed for only the magic-angle twist configuration\cite{Po2018, Yuan2018, Koshino2018}, 
typically ignoring atomic relaxation.

Here, we present an \textit{ab initio} $k \cdot p$ perturbation 
continuum model for tBLG which accurately accounts for the effects of atomic relaxation.
Our model reproduces the results of DFT-quality tight-binding hamiltonians
but at a smaller computational cost and, more importantly, it applies to 
all twist angles near the magic-angle value.
Such a single-particle model is a prerequisite for physically meaningful prediction of correlation effects, as the presence of unphysical features in the single-particle band structure causes uncontrolled error in many-body calculations.
We draw new conclusions on the low-energy electronic states at small twist angles, including the interesting 
result that there are no additional vanishings of the Fermi velocity in the range of the previously expected second and third magic-angles.
For reference, we compare our continuum model to the seminal and 
widely employed $k \cdot p$ model of Bistritzer and MacDonald\cite{Bistritzer2011}
(BMD in the following), and we adopt 
their dimensionless parameter $\alpha = \omega / v_F k_\theta$ for describing 
the twist-angle $\theta$, where $v_F$ is the Fermi velocity, 
$k_{\theta}$ is the wave-vector set by the moir\'{e} length scale
and $\omega$ is their effective interlayer coupling strength ($0.11$ eV).

\begin{figure*}[!htb]
  \centering
  \includegraphics[width=1\textwidth]{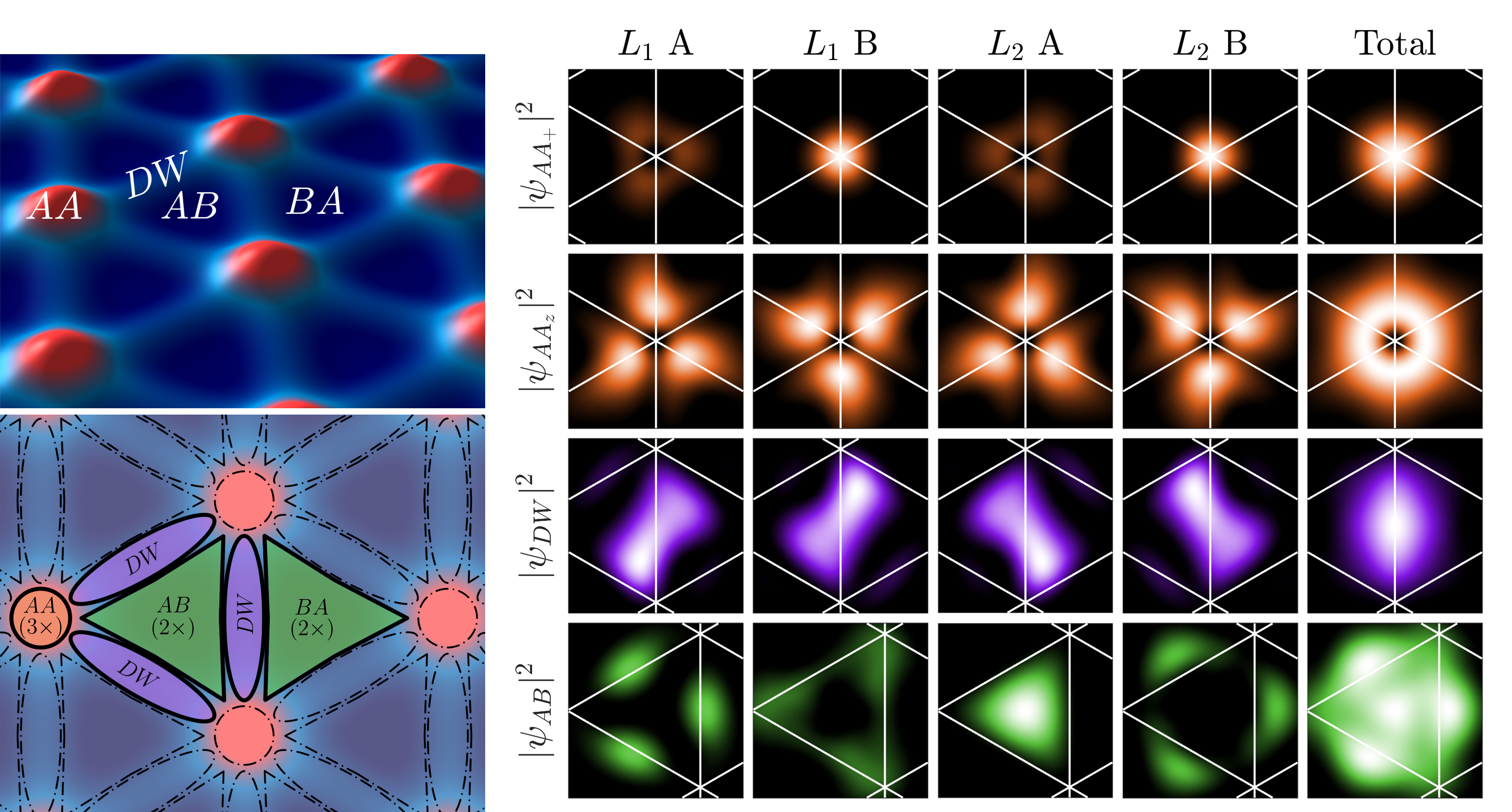}
  \caption{
  	Left: Structure of relaxed tBLG at $\theta = 0.9^\circ$ with exaggerated vertical relaxation (top).
	$AA$, $AB$ and $BA$ stackings, and domain walls ($DW$) are labeled along with
	a schematic representation (bottom) of the 10 orbitals 
per unit cell of the moir\'{e} pattern required to describe 
the low-energy electronic states:
	3 at the $AA$ region, 1 at each of the 3 $DW$ regions, and 2 at each of the 
$AB$ and $BA$ regions.
	Right: Wavefunction magnitudes, $|\psi_{l}|^2, l=AA_{\pm}, AA_{z}, DW, AB/BA$,
	 of the 10-band model, at $\theta = 0.9^\circ$,
projected in the two layers ($L_1$ and $L_2$) and the sublattices $A$ and $B$ 
of each layer; the total (far-right column) is the sum of all layer and sublattice contributions (see SM for additional discussion).
The underlying moir\'e supercell lattice is given by the thin white lines.
	}
  \label{fig:geom}
\end{figure*}

Within $k \cdot p$ perturbation theory, the set of Bloch states 
of the two graphene layers is augmented 
by the addition of interlayer couplings due to the twist-angle 
induced Umklapp scattering process.
As the low-energy electronic structure of tBLG 
is dominated by a pair of Dirac cones, 
the momentum expansion can be carried out about one copy of the cone at a valley K point.
Taking also into account spin degeneracy, 
each band represents four electronic states in a real system
\cite{ValleyNote}.
Here we introduce an expanded \textit{ab initio} $k \cdot p$ model which gives a more complete physical picture of the tBLG system.
Our model has three new key ingredients:\\
(1)  relaxation of the bilayer system \cite{Carr2018relax}, including the out-of-plane 
relaxation of different regions as well as the
in-plane strain corrections to the Hamiltonian of the individual monolayers;\\
(2) terms beyond the first shell of couplings in the $k \cdot p$ continuum model, 
which are necessary to capture the changes in stacking order at small angles;\\
(3) inclusion of $k$-dependent terms, which 
allow the $k \cdot p$ model to reproduce more accurately 
the particle-hole asymmetry of realistic \textit{ab initio} bandstructures.\\
The $k \cdot p$ terms are directly computed from an \textit{ab initio} tight-binding Hamiltonian model \cite{Fang2016, Fang2018, Carr2018pressure} for supercells spanning 
the twist-angle range $0.18^\circ \leq \theta \leq 6^\circ$.
These terms have smooth dependence on $\theta$, allowing for interpolation between 
the specific twist angles that correspond to finite supercells, 
to generate a model valid for any desired angle in that range.
We relegate the detailed description of the extended Hamiltonian and the 
procedure for obtaining the relevant terms of the continuum model to 
a companion paper \cite{FangPrep}.

Our continuum model affords a natural interpretation of the electronic structure 
of tBLG at small twist angles, which is derived directly from the atomic relaxation 
so we describe this aspect first.
For twist-angle $\theta$ smaller than a critical value 
$\theta_c \approx 1^\circ$, the local atomic structure near 
the $AA$ and $AB$ stackings of the two layers becomes {\em independent} of $\theta$.
This creates a pattern of small circular domains of $AA$ stacking 
and large triangular domains of $AB/BA$ stackings.
Domain walls ($DW$) of intermediate stacking separate the $AB$ and $BA$ domains
and connect the $AA$ regions.
This creates local electronic environments which are locked-in 
with respect to changing twist-angle for $\theta < \theta_c$, where 
the tBLG system consists of a few fixed elements\cite{Nam2017, Zhang2018, Carr2018relax},
and only their length scale changes for decreasing twist angle.
These elements are: the $AA$ regions which have a \textit{local} twist of $\theta_{AA}=1.7^\circ$, 
which is \textit{independent} of the overall twist angle $\theta$
between the two layers, the $AB$ and $BA$ regions with negligible local
twist.
Moreover, the diameter of the $AA$ regions and the width of the $DW$ regions
are approximately equal and remain unchanged for $\theta < \theta_c$\cite{Zhang2018}. 
These features are shown in Fig. \ref{fig:geom} for $\theta = 0.9^\circ$.

The relaxation in tBLG is described by two simultaneous effects. 
In-plane relaxation decreases the area of the high stacking energy $AA$ region
while it increases that of low stacking energy $AB/BA$ regions.
Out-of-plane relaxation causes corrugation, 
increasing the vertical separation between the $AA$ regions from the 
equilibrium distance in $AB$ stacking of 3.35 \AA\ to 3.59 \AA, 
a substantial change ($>7$\%). 
The reduction in the size of $AA$ stacking can be understood as a minimization of 
planar stress energy and stacking energies, 
and has been modeled through various methods \cite{Dai2016, Nam2017, Zhang2018, Carr2018relax} leading to a relaxed pattern in 
agreement with experimental results \cite{Alden2013, Yoo2018}.
The role of vertical relaxation 
in experimental devices is less understood, 
as only free-standing tBLG has been modeled.
Experimental tBLG devices are typically 
encapsulated in hexagonal Boron Nitride, so the actual corrugation  
may be reduced compared to the free-standing case.
To take this into account, we consider two limits of the vertical relaxation: 
a ``Full'' relaxation model (free-standing bilayer result)
and  a ``Flat'' model with constant interlayer distance equal 
to the average of $AA$ and $AB$ interlayer distances (3.47 \AA).
The magic angle predicted by the fully relaxed model, $\theta_c \approx 1.0^\circ$, is closer 
to the angles where correlated phenomena are observed \cite{Cao2018mott, Cao2018sc, Yankowitz2019}.

\begin{figure}
  \centering
  \includegraphics[width=0.5\textwidth]{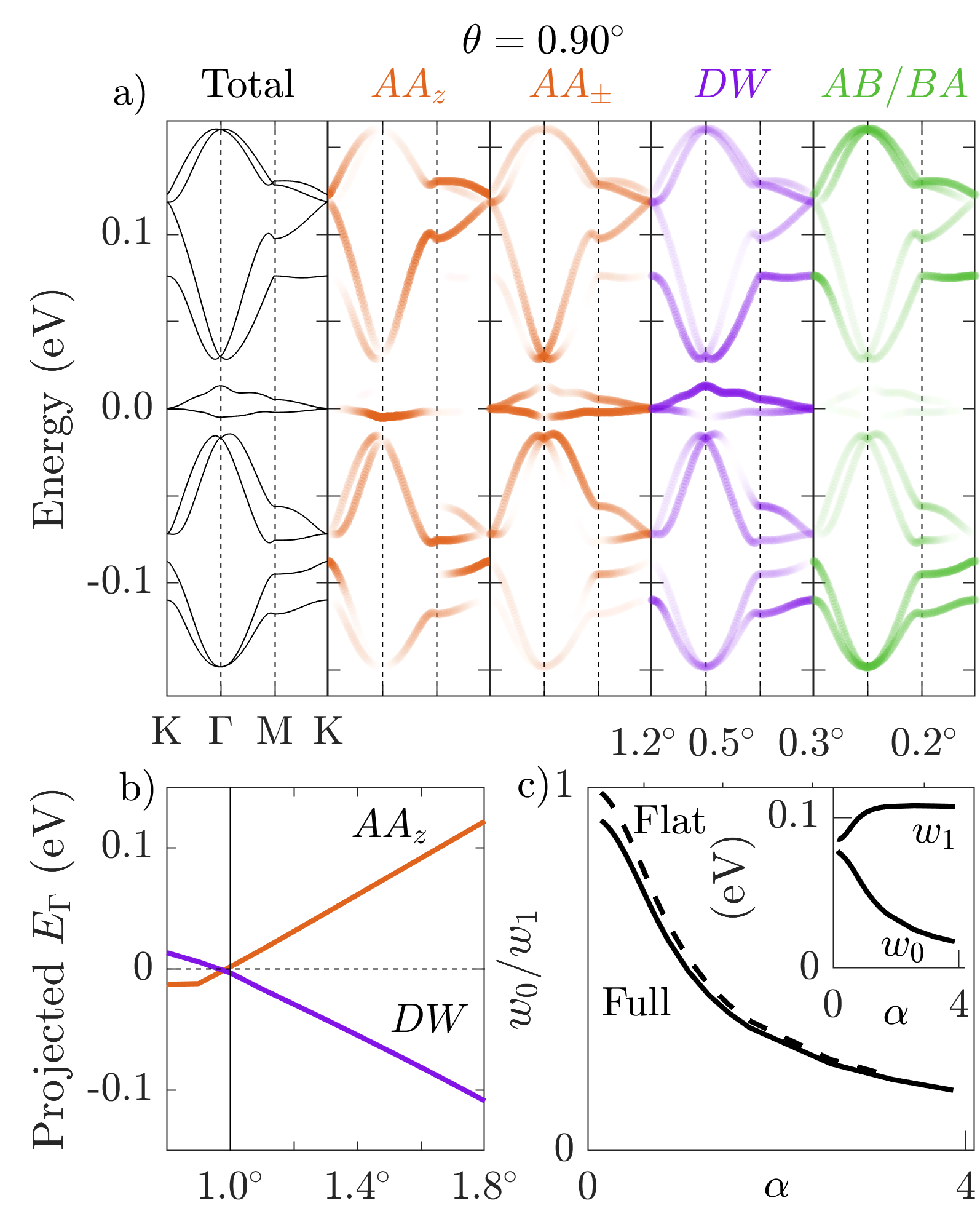}
  \caption{\textbf{(a)} Orbital character of the bands in the reduced 
10-band $k \cdot p$ model 
at $\theta = 0.90^\circ < \theta_c$.  
  \textbf{(b)} Energy of the flat-bands at the $\Gamma$ point ($E_\Gamma$) as function of $\theta$.
  One band is always $AA_z$ character and one is $DW$ character.
  The magic-angle regime is coincident with a change in the band character ordering.
  \textbf{(c)} Dependence of the interlayer $k \cdot p$ coupling terms on the twist angle $\theta$ from $6^\circ$ to $0.18^\circ$. 
  The inset gives the value of the individual terms with full relaxation, and the main panel gives their ratio for both flat and full relaxation.  
}
  \label{fig:10_band}
\end{figure}

The low-energy electronic states are directly associated with and derived from 
the presence of the relaxation-induced structural elements described earlier.
Since the discovery of correlated phases in tBLG, many simplified $n$-band models have been proposed for the flat-bands, usually based on localized functions of the BMD model.
One such minimal model consists of 10 bands \cite{Po2018}, and we argue that it can accurately capture the electronic effects of the different stacking regions that emerge after relaxation.
This model comprises three orbitals on a triangular lattice formed by the $AA$ sites,
one of $p_z$-like character ($AA_z$) and two of 
$(p_x \pm i p_y)$-like character ($AA_{\pm}$),  
three orbitals on a Kagome lattice formed by the domain walls,
and four orbitals on a honeycomb lattice,
two for each of the $AB$ and $BA$ domains.
The full details of the $10$-band tight-binding Hamiltonian are provided in the supplementary materials.

To compare our \textit{ab initio} $k \cdot p$ results to the $10$-band model, 
we project non-orthogonal wavefunctions that satisfy the symmetry conditions, shown in  Fig. \ref{fig:geom},
from band structure calculations.
The form of these wavefunctions is not sensitive to the twist-angle, 
and is robust for twist angles within $\pm 0.2^\circ$ of the magic angle.
We note that the $z$ and $\pm$ indexing of the $AA$ orbitals describe their symmetry properties over the moir\'e supercell, \textit{not} their composition in terms of atomic-scale C $p_z$ orbitals.
We also fit the parameters of the $10$-band tight-binding model for $\theta \in [0.8^\circ, 1.8^\circ]$, 
to reproduce the bands produced by our continuum model (see SM).
The flat bands near the magic angle have $AA$ and $DW$ character 
(see Fig. \ref{fig:10_band}a), showing that the coupling between these states 
is a necessary ingredient of the model if it is to capture the electronic structure
as a function of twist angle.
In particular, the orbital character of the electron and hole bands at $\Gamma$
flips as one reduces the twisting angle: 
the hole band has $DW$ character for $\theta>\theta_c$ 
and switches to $AA_{z}$ character for $\theta<\theta_c$,
while the electron band has the reverse character.
As the $AA_{z}$ and $DW$ orbitals have opposite $xy$-plane mirror symmetry eigenvalues ($-1$ and $+1$, respectively), the magic-angle represents a symmetry-protected band inversion.

Two other important parameters in the $k \cdot p$ model are the effective interlayer coupling between orbitals of the same sublattice label, $A \to A$ or $B \to B$, and that between
orbitals of different labels, $A \to B$ or $B \to A$.
These nearest-neighbor interlayer couplings have been labeled $w_i, i=0,1$ 
in previous studies and have a simple geometric interpretation:
$w_0$ is the interlayer electronic coupling at the $AA$ sites and $w_1$ is the  
coupling at $AB/BA$ sites, averaged over the entire moir\'e cell.
The values of these $w_i$ parameters depend strongly on the twist angle $\theta$.
As the lattice relaxes, the relative size of the $AA$ regions is greatly reduced 
while that of the $AB/BA$ regions is increased, 
causing a reduction in the value of $w_0$ and a modest increase in the value of $w_1$.
This dependence is shown in Fig. \ref{fig:10_band}c for the Full and the Flat relaxation models.
The overall $\theta$ dependence of the ratio $w_0/w_1$ is not sensitive to the 
relaxed height assumption.
The Flat model has a larger ratio as the Full relaxation assumption moves the $AB/BA$ 
sites closer together (increasing their coupling and the $w_1$ value) 
while moving the $AA$ sites farther apart (reducing their coupling and the $w_0$ value).

\begin{figure}
  \centering
  \includegraphics[width=.5\textwidth]{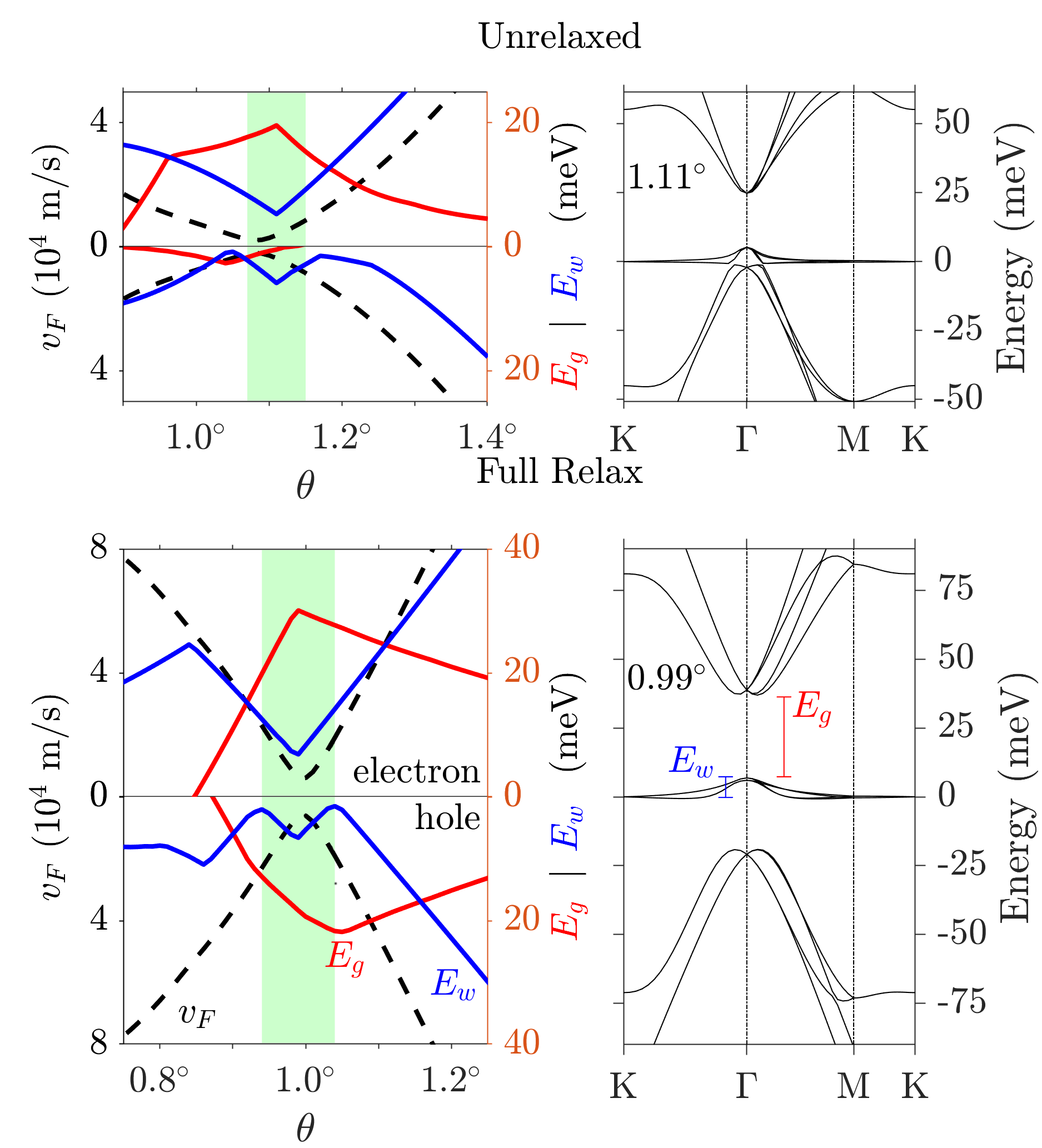}
  \caption{
  Left panels: Features of the flat bands near the magic-angle for models with 
  or without atomic relaxation: the Fermi velocity $v_F$ (dashed black line, left axis), 
  band gap $E_g$ (red lines, right axis), and bandwidth $E_w$ (blue lines, right axis) 
  for the electron and hole states.   
  In-plane relaxation creates a more well defined magic angle regime (green shaded region) in all three features.
  Right panels: corresponding bandstructures in the magic-angle regime.
  }
  \label{fig:magic_angle_compare}
\end{figure}

To elucidate the salient features of the single-particle model, 
we study three related indicators of the flat-band phenomenon as a function of $\theta$: the Fermi velocity ($v_F$), the bandwidth ($E_w$), and the band gap ($E_g$).
These are shown in Fig \ref{fig:magic_angle_compare}.
All three are calculated for both the electron and the hole sides of the flat-band manifold.
The model {\em without} relaxation shows large discrepancies between the extrema of the 
Fermi velocity, gap, and bandwidth, and the electron and hole features have little in common.
The two models (Flat and Full) that include relaxation show more regular dependence on $\theta$ 
and closer correspondence between the electron and hole  bands.
The bandwidth for the hole band is always smaller than that of the electron band, 
and the hole band achieves its minimum twice.
In general, $v_F = 0$ does not coincide with bandwidth minima.
We thus draw the important conclusion that 
\textit{the magic-angle is not a single value, but rather a range of $\approx 0.1^\circ$
which spans the extrema in these key features}.
In particular, even if an experimental device has a variation in twisting angle over a probed region,
if that variation is $\approx 0.1^\circ$ the flat-band models may still be reliable enough to explain correlation effects.
This range for the Full relaxed model is $\theta \in [0.95^\circ, 1.05^\circ]$ 
and $\theta \in [0.80^\circ,0.90^\circ]$ for the Flat model.
The bandstructures for both models are similar after accounting for this offset in $\theta$.

An interesting behavior of the Full relaxed model occurs at the center of the magic-angle regime: although the Dirac cone still has symmetric dispersion near the K-point, 
the hole band dispersion is such that near the $\Gamma$ point its energy energy is {\em higher} 
than the Fermi level (see Fig. \ref{fig:magic_angle_compare}).
Thus the charge neutrality point does {\em not} occur at the Dirac point energy.
This effect persists in all of our \textit{ab initio} $k \cdot p$ models 
(even without relaxation), 
and is a behavior that can be observed 
in other tight-binding models in the literature\cite{Sboychakov2015, Nam2017, Angeli2018}.
Assuming the bands of tBLG are not perfectly particle-hole symmetric, and that the flat-band regime is defined by a protected $\theta$-tuned band inversion, such a feature is unavoidable.
For transport measurement, this behavior would result in 
a range of $0.1^\circ$ in twist angle where the charge neutrality point 
of the flat bands does not align with the Dirac-point of the moir\'e superlattice, 
as well as a reduction in the resistivity at the Dirac-point energy due to these other bands near $\Gamma$.
Thus if a clean Dirac-point transport signature is used to assess experimental device quality, this angle-range will be difficult to observe.

\begin{figure}
  \centering
  \includegraphics[width=.5\textwidth]{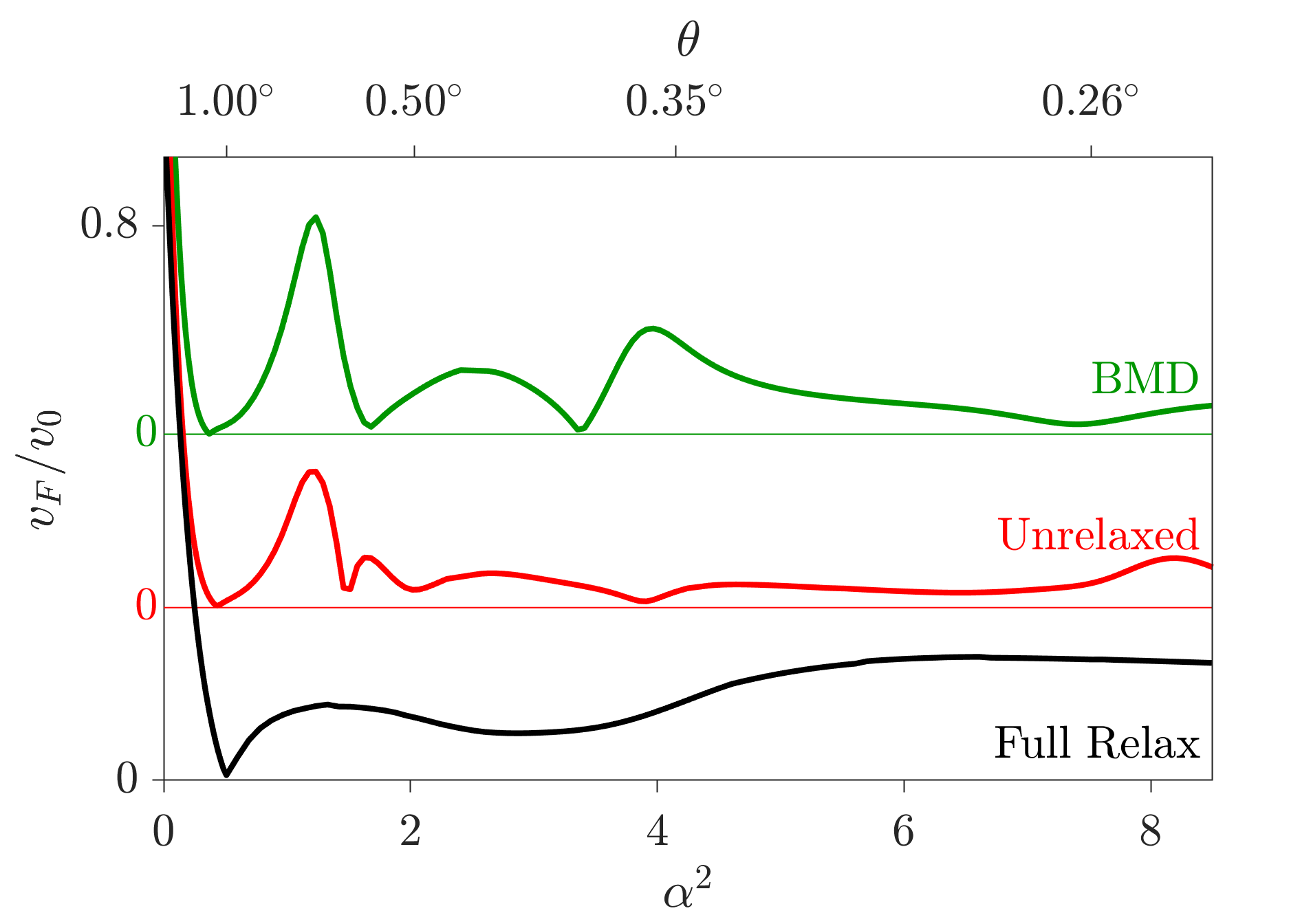}
  \caption{Normalized Fermi velocity as a function of $\alpha^2 \propto 1/\theta^2$ for the 
  BMD and the {\em ab initio} $k \cdot p$  models without relaxation (Unrelaxed) 
  and with atomic relaxation (Full relaxed).}
  \label{fig:vf_compare}
\end{figure}

Another important result of our calculations including atomic relaxation in tBLG is 
the suppression of the second magic-angle twist, 
defined as a smaller twist angle at which $v_F = 0$ \cite{Bistritzer2011}.
In Fig. \ref{fig:vf_compare} we show 
the Fermi velocity as predicted from the BMD model 
and from our unrelaxed and fully relaxed \textit{ab initio} $k \cdot p$ models. 
Although our unrelaxed model shows similar behavior to the BMD model with 
a second magic angle occurring near $\theta = 0.5^\circ$, 
the inclusion of atomic relaxation removes this feature in near $0.5^\circ$.
As the lattice relaxation in tBLG becomes increasingly sharp on the moir\'e length scale as 
the twist angle decreases \cite{Dai2016, Nam2017, Zhang2018, Carr2018relax, Yoo2018}, these sharper features
in the relaxation introduce additional important couplings in the $k \cdot p$ model at larger momenta.
Thus to accurately model the electronic structure of tBLG below $1^\circ$ our inclusion of the higher-order $k \cdot p$ couplings terms is necessary.

In conclusion, we have presented a $k \cdot p$ expansion of the low-energy electronic states of tBLG that can be extended to arbitrary order in pertubation theory.
This exact continuum model facilitates a better understanding of the single-particle features of tBLG's flat bands, and provides a solid foundation on which to build correlated models.
We have made this model publicly available in MATLAB, C++, and Python at \url{https://github.com/stcarr/kp_tblg}.

\begin{acknowledgments}
We thank Daniel Massatt, Hoi Chun Po, Alex Kruchkov, Grigory Tarnopolskiy, Pablo Jarillo-Herrero, Hyobin Yoo, Rebecca Engelke, and Philip Kim for useful discussions. This work was supported by ARO MURI Award W911NF-14-0247 and by the STC Center for Integrated Quantum Materials, NSF Grant No. DMR-1231319. The computations in this paper were run on the Odyssey cluster supported by the FAS Division of Science, Research Computing Group at Harvard University.
\end{acknowledgments}

\bibliographystyle{apsrev4-1}

\bibliography{relaxed_kp_tblg}

\begin{thebibliography}{31}
\bibitem [{Val()}]{ValleyNote}%
  \BibitemOpen
  \href@noop {} {\bibinfo  {journal} {Here we consider only a single
  (monolayer) valley, but the contribution of the second valley can be included
  by sampling the $C_2$ transformation of the original path in momentum space.
  The two valleys give slightly different band-structures along the $\Gamma$ to
  $M$ line but are identical along other high-symmetry lines.}\ }\BibitemShut
  {NoStop}%
\end{thebibliography}

\end{document}


\preprint{APS/123-QED}

\title{Supplementary Material: Minimal model for low-energy electronic states of twisted bilayer graphene}
\author{Stephen Carr}
\affiliation{Department of Physics, Harvard University, Cambridge, Massachusetts 02138, USA}
\author{Shiang Fang}
\affiliation{Department of Physics, Harvard University, Cambridge, Massachusetts 02138, USA}
\author{Ziyan Zhu}
\affiliation{Department of Physics, Harvard University, Cambridge, Massachusetts 02138, USA}
\author{Efthimios Kaxiras}
\affiliation{Department of Physics, Harvard University, Cambridge, Massachusetts 02138, USA}
\affiliation{John A. Paulson School of Engineering and Applied Sciences, Harvard University, Cambridge, Massachusetts 02138, USA}

\maketitle

\section{Relaxation of twisted bilayer graphene}

The rigid twisted bilayer is subject to the atomic relaxations to minimize the total energy of the crystal.
Here we assume $D_6$ rotational symmetry at an $AA$ stacking site (e.g. the rotation axis is through the center of a hexagon, not a Carbon site).
Regions with different local atomic registry differ in their stacking energy.
We use a generalized stacking fault energy (GSFE) functional from DFT calculations \cite{Zhou2015} to capture this registry-dependent energy.
The conventional $AB/BA$ (Bernal) stacking are of lowest energy, and so relaxed domains of locally $AB/BA$ stacking are favored in the twisted bilayer graphene crystal at angles below $2^\circ$.
Regions of $AA$ stacking spots are reduced with domain boundary line formation in between neighboring $AB/BA$ regions.
Another feature is the puckered out-of-plane crystal relaxations.
The vertical layer separation is shortest (longest) for the locally $AB/BA$ ($AA$) stacking region. 
The above qualitative description of the relaxed twisted bilayer graphene crystal is substantiated by \textit{ab initio} calculations and continuum models, and for further details we refer to Ref. \onlinecite{Carr2018relax}.

The structural deformation can be described by an in-plane shift vector $\vec{u}^i(\vec{r})=(u_x^i(\vec{r}),u_y^i(\vec{r}))$ and an out-of-plane component $h^i(\vec{r})$.
This causes lattice relaxation by mapping unrelaxed positions $\vec{r}$ in the $i$-th layer of the rigid bilayer to $\vec{r}+\vec{u}^i(\vec{r})+h^i(\vec{r}) \hat{z}$.
The relaxation pattern respects the three-fold rotation symmetry and mirror symmetry of the bilayer, and the functional form can be Fourier expanded:

\begin{equation}
\vec{u}^i(\vec{r})=-i\sum_{\vec{q}} u_{\vec{q}}^i e^{i\vec{q} \cdot \vec{r}}, h^i(\vec{r})=h_0+\sum_{\vec{q}} h_{\vec{q}} e^{i\vec{q} \cdot \vec{r}}
\end{equation}

Symmetry requires $u_{\vec{q}'}^1 = R_{60^{\circ}} u_{\vec{q}}^1$ and $h_{\vec{q}'}=h_{\vec{q}}$ when $\vec{q}'$ is $60^{\circ}$ rotated from $\vec{q}$ and $R_{60^{\circ}}$ the $60^{\circ}$ rotation matrix for the vector.
The $\vec{q}$'s are sums of the reciprocal lattice parameters, $\vec{q} = m\vec{G}_1 + n \vec{G}_2$ for integers $m,n$ and reciprocal lattice vectors $G_i$.
In practice, we use up to $m,n = 20$ to model the relaxation accurately down to $\theta = 0.18^\circ$, but for $\theta \geq 1^\circ$ only the first two ($m,n = 2$) components are usually needed.

\pagebreak

\section{Continuum Hamiltonian for relaxed tBLG} 

Our low-energy effective $k \cdot p$ Hamiltonian for relaxed twisted bilayer graphene (tBLG) has the following form:

\begin{equation}
\tilde{H}_{\rm K}=\begin{bmatrix}
H^1_D(\vec{k}) +A_1(\vec{r}) +V_1(\vec{r})& \tilde{T}^\dagger(\vec{r}) + \big\{M^{\dagger}_+(\vec{r}),\hat{k}_-  \big\} +\big\{ M^{\dagger}_-(\vec{r}) ,\hat{k}_+ \big\} \\
\tilde{T}(\vec{r})+\big\{ M_+(\vec{r}) ,\hat{k}_+  \big\} +\big\{ M_-(\vec{r}) ,\hat{k}_- \big\} & H^2_D(\vec{k}) +A_2(\vec{r}) +V_2(\vec{r})
\end{bmatrix}
\end{equation} 

\begin{itemize}
    \item $H^i_D(\vec{k})$ is the Dirac Hamiltonian for each individual graphene monolayer layer.
    \item $V_i(\vec{r})$ terms are the external potential we can introduce to each layer. These could include the electric gating potential, sublattice mass terms and the modulated electric potential from doping and charge redistribution.
    \item $A_i(\vec{r})$ is the in-plane pseudo-gauge field coupled to the Dirac electron. These terms are generated from the geometric deformation and strain for each layer. When expanded in Fourier components with supercell reciprocal lattice vectors, it can be written as $A_i(\vec{r}) = \sum_{\vec{p}_i} A^i_{\vec{p}_i} e^{i \vec{p}_i \cdot \vec{r}}$.
    \item The first part of the inter-layer coupling term $\tilde{T}(\vec{r})$ gives the scattering terms as in the original BMD model \cite{Bistritzer2011}. However, we generalize the expansion to include higher-order terms as $\tilde{T}(\vec{r}) = \sum_{\vec{q}_i} \tilde{T}_{\vec{q}_i} e^{i \vec{q}_i \cdot \vec{r}}$.
    \item The remaining part of the inter-layer coupling terms, with $\hat{k}_\pm = \hat{k}_x \pm i \hat{k}_y$, are the momentum dependent scattering terms. They are relevant for the particle-hole asymmetric features of the tight-binding band structures. The anti-commutator notation is used to symmetrize the non-commuting operators $\hat{r}$ and $\hat{k}$.
    \item The coupling constants here are investigated numerically with twist angle dependence. They are derived from the projection of supercell calculations with relaxed geometry obtained from elastic theory.
\end{itemize}

\pagebreak

\section{Validation of band structures}

Here we compare of bandstructures of our full tight-binding model, our $k \cdot p$ model, and the reduced $10$-band model \cite{Po2018} across four representative angles.
The $k \cdot p$ model perfectly recreates the low-energy tight-binding band structures if we sample over both valleys of the monolayers (their bands only differ here on the $\Gamma$ to $M$ line).
The 10-band model for each angle is obtained by first computing the $k \cdot p$ model, and then optimizing the 10-band model's 18 parameters to minimize

$$
\textrm{Err} = \sum_{b = 1}^{10} w_b \sqrt{ \sum_{\vec{k}} ( E_b(\vec{k}) - \epsilon_b(\vec{k}) )^2 }
$$

where $E_b$ are the eigenvalues of the 10-band model for band $b$ and $\epsilon_b$ are the pre-computed eigenvalues for the $k \cdot p$ model.
As we want the model to be most accurate near the Dirac cone energy ($E = 0$), we use $w_b$ to weigh the central bands higher than the outer bands during the optimization procedure.

\begin{figure}[b!]
  \centering
  \includegraphics[width=0.9\textwidth]{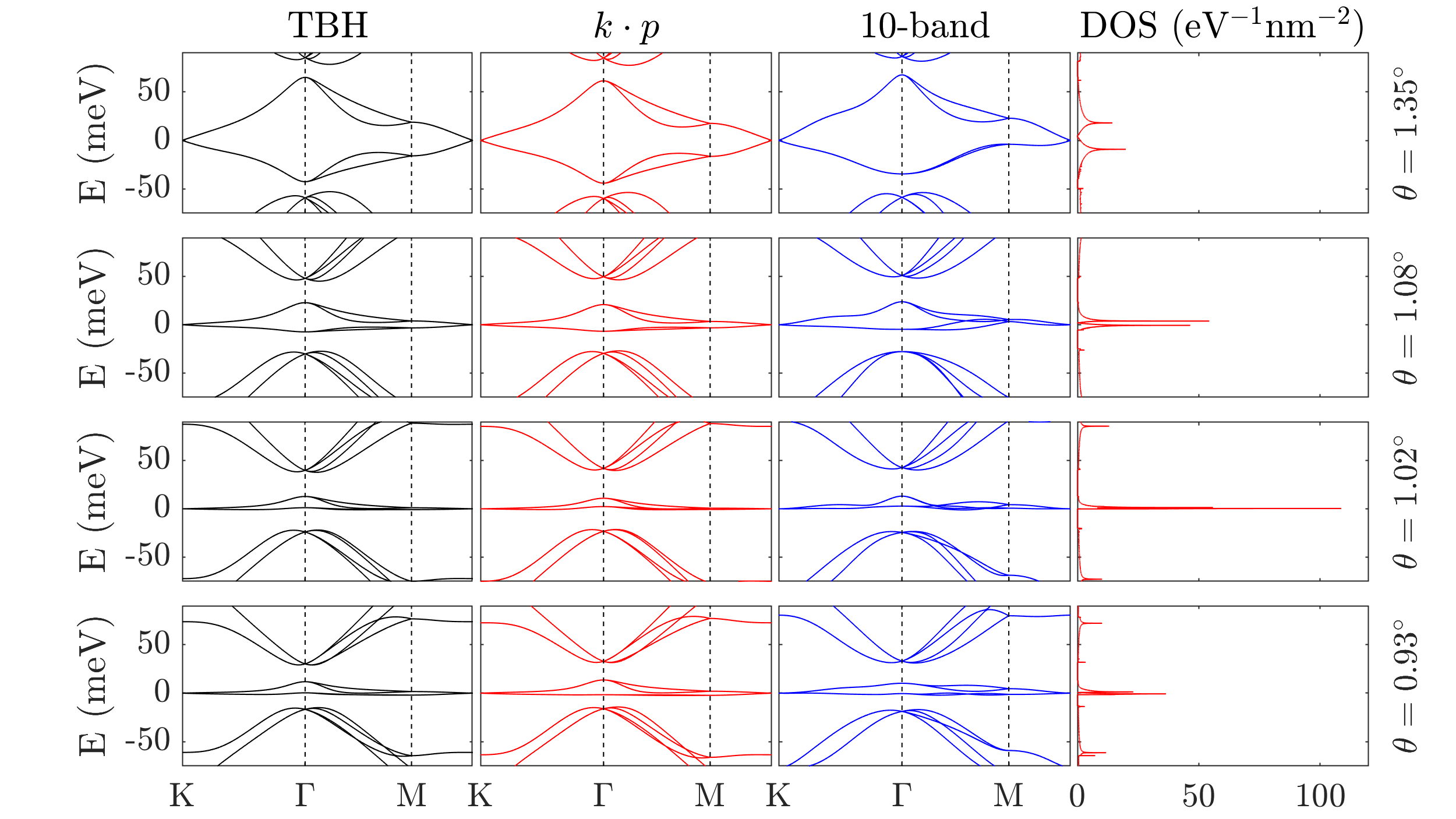}
  \caption{
  	Band structures for fully relaxed twisted bilayer graphene for a full tight-binding Hamitlonian, our \textit{ab initio} $k \cdot p$ model, and a fitted 10-band model \cite{Po2018}.
  	Four angles are checked, one well above the magic angle ($1.35^\circ$), one slightly above ($1.08^\circ$), one very close ($1.02^\circ$) and one slightly below ($0.93^\circ$).
  	The Density of Stats (DOS) is calculated from a $40 \times 40$ sampling of the supercell Brillioun zone within the $k \cdot p$ model.
	}
  \label{fig:projections_full}
\end{figure}

\pagebreak

\section{10-band Hamiltonian}

Following Ref. \onlinecite{Po2018}, we present the 10-band Hamiltonian.
Our labeling of the 10 bands is connected to Ref. \onlinecite{Po2018} in the following way: $AA_z \to p_z$, $AA_\pm \to p_\pm$, $DW \to \kappa$, and $AB/BA \to \eta$.
The Hamiltonian is written in the momentum basis for simplicity, but each term represents a bond between two effective atomic orbitals in a realspace Hamiltonian.
To simplify the notation, the following symbols are used to represent phases arising from bonds crossing the primitive unit-cell of the moir\'e cell or rotational symmetry constraints:

$$
\phi_{l m} \equiv e^{-i \mathbf{k} \cdot (l \mathbf{a}_1 + m \mathbf{a}_2)}, w \equiv e^{i 2 \pi / 3}, \zeta \equiv e^{i 2 \pi / 6}.
$$

%


The 10-band Hamiltonian has diagonal blocks made of 4 orbital lattices: $AA_z$ ($1 \times 1$), $AA_\pm$ ($ 2 \times 2 $), $DW$ ($3 \times 3$), and $AB/BA$ ($4 \times 4$).
The off-diagonal blocks, labeled as $C$ matrices, represent couplings between different orbital lattices.

$$
H_\mathbf{k} = 
\begin{pmatrix}
H_{AA_z} + \mu_{AA_z} & C^\dag_{AA_\pm, AA_z} & 0 & C^\dag_{AB/BA,AA_z} \\
C^\dag_{AA_\pm, AA_z} & H_{AA_\pm} + \mu_{AA_\pm} & C^\dag_{DW,AA_\pm} & C^\dag_{AB/BA,AA_\pm} \\
0 & C_{DW,AA_{\pm}} & H_{DW} + \mu_{DW} & C^\dag_{AB/BA, DW} \\
C_{AB/BA,AA_z} & C_{AB/BA,AA_\pm} & C_{AB/BA,DW} & H_{AB/BA} 
\end{pmatrix}
$$

Note that the term $C_{DW, AA_z}$ is set to zero here but in principle can be non-zero.
the $\mu_i$ are multiples of the identity matrix which are the direct chemical potentials of the orbital lattices.
However, due to the couplings present in each lattice the energy level of the isolated system in the flat-band manifold is slightly different (and called $\delta_i$ in the main text). 
The value of $\delta_i$ are given in Table \ref{tab:chem_pots}.
The value of $\delta_{p_z}$ and $\delta_{\kappa}$ control the energies of the $AA_z$ and $DW$ portions of the flat bands at $\Gamma$, while $\delta_{\pm}$ controls the location of the Dirac cone touching point at $K$ from $AA_\pm$ ($p_x \pm i p_y$ type orbitals).

\begin{table}[!h]
\begin{tabular}{l|c|l}
 Param. & Orbital & Value \\
 \hline 
 $\delta_{p_z}$ 		& $AA_{z}$  	& $\mu_{p_z} + 6 t_{p_z}$ \\
 $\delta_{p_{\pm} }$ 	& $AA_{\pm}$  	& $\mu_{p_{\pm} } - 3 t_{p_{\pm} }$ \\
 $\delta_{\kappa}$ 		& $DW$  		& $\mu_{\kappa} + 4 (t_{\kappa} + t'_{\kappa})$ \\
  \end{tabular}
  \caption{Definition of three effective chemical potentials in the 10-band model \cite{Po2018} and the geometric orbitals they are associated with.}
\label{tab:chem_pots}
\end{table}

We next tabulate the form of the orbital lattices on the diagonal blocks of the Hamiltonian.
These generally depend on one parameter ($t_i$), or two parameters ($t^\pm_i$) where the $\pm$ indexing is specifically chosen to allow for the breaking of time-reversal symmetry.
The term $t'_\kappa$ is primed to remind that it is a second nearest-neighbor bond, as having only first nearest-neighbor bonding on the Kagome lattice introduces a flat-band unrelated to the magic-angle phenomenon.

$$
H_{AA_z} = t_{p_z} (\phi_{0 1} + \phi_{11} + \phi_{10} + h.c.)
$$

$$
H_{AA_\pm} = t_{p_\pm} (\phi_{01} + \phi_{11} + \phi_{10} + h.c.) 
\begin{pmatrix}
1 & 0 \\ 
0 & 1
\end{pmatrix}
+ 
\begin{pmatrix}
0 & C^\dag_{p_\pm p_\pm} \\
C_{p_\pm p_\pm} & 0
\end{pmatrix}
$$

$$
C_{p_\pm p_\pm} = t^+_{p_\pm p_\pm} (\phi_{01} + \phi_{\bar{1} \bar{1}} \omega + \phi_{10} \omega^*) + t^-_{p_\pm p_\pm} (\phi_{0 \bar{1}} + \phi_{11} \omega + \phi_{\bar{1}0} \omega^*)
$$

$$
H_{DW} = t_\kappa
\begin{pmatrix}
0 & \phi_{\bar{1}0} & 1 \\
1 & 0 & \phi_{0\bar{1}} \\
\phi_{11} & 1 & 0
\end{pmatrix}
+ t'_\kappa
\begin{pmatrix}
0 & \phi_{\bar{1}\bar{1}} & \phi_{\bar{1}0} \\
\phi_{0\bar{1}} & 0 & \phi_{10} \\
\phi_{01} & \phi_{11} & 0 
\end{pmatrix} + h.c.
$$

$$
H_{AB/BA} = t_\eta
\begin{pmatrix}
0 & e^{i \phi_\eta} (1  + \phi_{0 \bar{1}} + \phi_{10}) \\ 
e^{-i \phi_\eta} (1  + \phi_{01} + \phi_{\bar{1}0}) & 0 
\end{pmatrix}
\otimes 1_{2 \times 2}
$$

Finally, we tabulate the form of the inter-orbital hoppings.
The formulae for $C_{AB/BA, i}$ depend on four real numbers $(a,b,c,d)$ which represent how the $AB/BA$ wave-function overlaps the $AA$ and $DW$ orbitals (see Ref. \onlinecite{Po2018} for more details).

$$
C_{AA_\pm, AA_z} = i t^+_{p_\pm p_z} 
\begin{pmatrix}
\phi_{01} + \phi_{\bar{1}\bar{1}} \omega + \phi_{10} \omega^* \\
-(\phi_{0\bar{1}} + \phi_{11} \omega^* + \phi_{\bar{1}0} \omega)
\end{pmatrix}
- i t^-_{p_\pm p_z}
\begin{pmatrix}
\phi_{0\bar{1}} + \phi_{11} \omega + \phi_{\bar{1}0} \omega^* \\
-(\phi_{01} + \phi_{\bar{1}\bar{1}} \omega^* + \phi_{10} \omega)
\end{pmatrix}
$$

$$
C_{DW, AA_\pm} = t^+_{\kappa p_\pm}
\begin{pmatrix}
\phi_{\bar{1}0} & \phi_{\bar{1}\bar{1}} \\
\phi_{\bar{1}\bar{1}} \omega^* & \omega \\
\omega & \phi_{\bar{1}0} \omega^*
\end{pmatrix}
- t^-_{\kappa p_\pm}
\begin{pmatrix}
\phi_{\bar{1}\bar{1}} & \phi_{\bar{1}0} \\
\omega^* & \phi_{\bar{1}\bar{1}} \omega \\
\phi_{\bar{1}0} \omega & \omega^*
\end{pmatrix}
$$

$$
C_{AB/BA, AA_z} = a
\begin{pmatrix}
-(\omega^* + \phi_{\bar{1}\bar{1}} \omega + \phi_{0 \bar{1}}) \zeta \\
(1 + \phi_{\bar{1}\bar{1}} \omega + \phi_{0 \bar{1}} \omega^*) \zeta \\
(\omega^* + \phi_{\bar{1}0} \omega^* + \phi_{\bar{1}\bar{1}}) \zeta^* \\
-(\omega^* + \phi_{\bar{1}0} + \phi_{\bar{1}\bar{1}} \omega) \zeta^*
\end{pmatrix}
$$

$$
C_{AB/BA,AA_\pm} = 
\begin{pmatrix}
b(\omega + \phi_{\bar{1}\bar{1}} \omega^* + \phi_{0\bar{1}}) \zeta^* & c(1 + \phi_{\bar{1}\bar{1}} + \phi_{0\bar{1}}) \\
c(1 + \phi_{\bar{1}\bar{1}} + \phi_{0\bar{1}}) & b(1 + \omega^* \phi_{\bar{1}\bar{1}} + \omega \phi_{0\bar{1}}) \zeta^* \\
b(\omega + \phi_{\bar{1}0} \omega^* + \phi_{\bar{1}\bar{1}}) \zeta & c(1 + \phi_{\bar{1}0} + \phi_{\bar{1}\bar{1}}) \\
c(1 + \phi_{\bar{1}0} + \phi_{\bar{1}\bar{1}}) & b(\omega + \phi_{\bar{1}0} + \phi_{\bar{1}\bar{1}} \omega^*) \zeta
\end{pmatrix}
$$

$$
C_{AB/BA, DW} = d
\begin{pmatrix}
i \phi_{10} & i\omega^* & i \phi_{0\bar{1}} \omega \\
i \phi_{10} & i\omega & i \phi_{0\bar{1}} \omega^* \\
-i & -i \omega^* & -i \omega \\
-i & -i \omega & -i \omega^*
\end{pmatrix}
$$

The 10-band Hamiltonian depends on 18 tunable parameters.
We list them in Table \ref{tab:10_band_params} with a short description and the value obtained when fitted to a $\theta = 0.90^\circ$ $k \cdot p$ band-structure.
In general, we find that this 10-band model with mostly nearest-neighbor coupling cannot perfectly recreate the \textit{ab initio} $k \cdot p$ flat-bands in the magic-angle regime.
It is likely that additional second or third nearest-neighbors are required to robustly recreate the dispersion of the flat-bands near the magic-angle.
In future work we aim to generate these terms directly from a Wannierization of the $k \cdot p$ model.

\begin{table}[!h]

\begin{tabular}{p{1.4cm}|p{2cm}|p{1.5cm}}
 \hline
 Param. & Internal nnbr. of & \ \ Value \\
 \hline 
 $t_{p_z}$ & \ \ \ \ $AA_z$ & \ -3.661 \\
 $t_{p_\pm}$ & \ \ \ \  $AA_\pm$ & \  -0.205 \\
 $t^+_{p_\pm p_\pm}$ & \ \ \ \  $AA_\pm$  & \ -2.260 \\
 $t^-_{p_\pm p_\pm}$ & \ \ \ \  $AA_\pm$  & \ -1.661 \\
 $t_\kappa$ & \ \ \ \  $DW$ & \ \ 8.989 \\
 $t'_\kappa$ & \ \ \ \  $DW$ & \ -6.634 \\
 $t_\eta$ & \ \ $AB/BA$ & \ \ 39.72 \\
 \hline
\end{tabular}
\quad
\begin{tabular}{p{1.4cm}|p{2.5cm}|p{1.5cm}}
 \hline
 Param. & Inter Orbital nnbr. from & \ \ Value \\
 \hline 
 $t^+_{p_\pm p_z}$ & $AA_\pm$ to $AA_z$ & \ \ 3.831 \\
 $t^-_{p_\pm p_z}$ & $AA_\pm$ to $AA_z$ & \ \ 1.149 \\
 $t^+_{\kappa p_\pm}$ & $DW$ to $AA_\pm$ & \ -7.956 \\
 $t^-_{\kappa p_\pm}$ & $DW$ to $AA_\pm$ & \ -5.678 \\
\hline 
\end{tabular}
\vfill
\begin{tabular}{p{1.4cm}|p{2cm}|p{1.5cm}}
 \hline
 Param. & Eff. Chem. Pot. of & \ \ Value \\
 \hline  
 $\delta_{p_z}$ &  \ \ \ \  $AA_z$  & \ -12.10 \\
 $\delta_{p_\pm}$ &  \ \ \ \  $AA_\pm$ & \ \ 6.467\\
 $\delta_\kappa$ &  \ \ \ \ $DW$  & \ \ 5.933\\
 \hline
 \end{tabular}
\quad
 \begin{tabular}{p{1.4cm}|p{2.5cm}|p{1.5cm}}
  \hline
 Param. & Inter Orbital from $AB/BA$ & \ \ Value \\
 \hline 
 $a$ &  \ \ \ \ to $AA_z$ & \ -24.69 \\
 $b$ & \ \ \ \ to $AA_\pm$ & \ -22.89 \\
 $c$ & \ \ \ \ to $AA_\pm$ & \ -10.51 \\
 $d$ & \ \ \ \ to $DW$ & \ \ 36.67 \\ 
 \hline
 \end{tabular}
  \caption{10-band model parameters and values for $\theta = 0.9^\circ$ given in meV.}
\label{tab:10_band_params}
\end{table}

\section{Projections of the 10-band model}

To obtain projections onto the 10-band model \cite{Po2018}, we begin with a trial wavefunction $\phi_0$ with appropriate layer and sublattice symmetry.
Next, we solve for $n$ of the low-energy eigenvectors $\psi^j_{\vec{k}}$ of our $k \cdot p$ Hamiltonian $H_{\vec{k}}$, where $\vec{k}$ spans a mesh sampling of the supercell Brilloun zone.
Note that $\psi^j_{\vec{k}}$ is supported on a basis of \textit{different} momenta, $\vec{q}$, which label the degrees of freedom of the  $k \cdot p$ Hamiltonian, $H_{\vec{k}}$.
We then compute the trial wavefunction's projection onto each $H_{\vec{k}}$,
$\phi^j_{\vec{k}} =  \sum_{\vec{q}} e^{i \vec{q} \cdot \vec{r}} \ket{\psi^j_{\vec{k}} (\vec{q})} \bra{\psi^j_{\vec{k}} (\vec{q})} \phi_0(\vec{r})$.
Finally, we reconstruct the real space projection via $\phi(\vec{r}) = \int d\vec{k} \sum_{\vec{q}} e^{i \vec{q} \cdot \vec{r} } \phi^j_{\vec{k}}$, where $\vec{q}$ are appropriate for $H_{\vec{k}}$.

In Fig. 1 of the main text, these projections are shown for a $k \cdot p$ model at $\theta = 0.90^\circ$.
Importantly, only one valley is considered in this projection, and we note that for a full physical tight-binding model, an additional 10 orbitals coming from the other monolayer valley must be included.
With this is mind, we discuss the orbitals themselves.
The $AA$ orbitals are named after $p$-type orbitals, but that only describes their angular momenta: the orbitals themselves do not look like conventional $p$-orbitals on the moir\'e supercell.
$\psi_{AA_+}$ sits on a triangular lattice with most of its density on the $B$ orbitals near the $AA$ spot.
To get $\psi_{AA_-}$ one performs a mirror-plane symmetry bisecting the axis of rotation, putting the density onto the $A$ orbitals near AA stacking.
$\psi_{AA_z}$ is also centered on the triangular lattice, but its density forms a three-lobed shape in each layer and sublattice index: it overlaps more strongly with the domain wall orbitals. 
$\psi_{DW}$ sits on one third of a Kagome lattice, and the other $2$ $DW$-like orbitals lie on the other $2$ Kagome lattice sites (centers of the white lines in Fig. 1 of the main text).
$\psi_{AB}$ has a partner orbital also centered on the $AB$ region, and can be thought of as a swapping of both the layer and sublattice index.
That is to say, $\psi_{AB}$ is primarily on sublattice $A$ of layer $2$, while its partner will be primarily on $B$ of layer $1$.
There are also two orbitals centered on the $BA$ region, with one primarily on sublattice $A$ of layer $1$ and the other on $B$ of layer $2$.
This projection technique will be replaced by a proper Wannierization of the \textit{ab initio} $k \cdot p$ model in the future, so we do not report numerical details of the projections.

\newpage

\section{On the accuracy of the model}

Although the model is based on \textit{ab initio} calculations, there may be shortcomings when comparing to experiment.
For future comparison, we list some of the ways the magic-angle regime can be ``shifted'' in our model, due to varying assumptions placed on the $k \cdot p$ parameters.
From DFT we obtain a graphene Fermi velocity of $v_F \approx 0.8 \times 10^6$ m/s, but changing the Fermi velocity proportionally changes the magic-angle.
Increasing $v_F$ by $10\%$ will \textit{decrease} the magic angle by $10\%$.
Similarly, increasing the interlayer coupling by $10\%$ will \textit{increase} the magic-angle by $10\%$ (this is the primary cause of the difference between the full and flat relaxed model).
However, both $v_F$ and the inter-layer coupling are directly proportional to ``nearest-neighbor'' tight-binding couplings (either intra-layer or inter-layer respectively).
We expect that improving the DFT calculation, by e.g. including the GW approximation, would increase both $v_F$ and the interlayer coupling, so it is not clear if such improvements would increase or decrease the magic-angle.
The straining of the graphene layer shows up as a pseudo-gauge field for the monolayer Hamiltonians of our $k \cdot p$ model, moving the location of the Dirac cones in momentum space.
Including this strain term generally \textit{decreases} the magic-angle by $10\%$, moving the magic angle regime from $1.1^\circ$ in the unrelaxed model to $1.0^\circ$ in the fully relaxed model, consistent with previous tight-binding modeling \cite{Nam2017}.
It is also possible that the bilayer's energy can be further optimized by allowing for shearing between the layers, and this may move the magic-angle regime by up to $5\%$ \cite{Lin2018}.

\bibliographystyle{apsrev4-1}

\bibliography{relaxed_kp_tblg}